\documentclass[twocolumn,amsmath,amssymb,prl,aps]{revtex4}

\begin{document}

\title{Solar neutrinos and the solar composition problem}

\author{Carlos Pe\~na-Garay\thanks{E-mail: penya@ific.uv.es}}
\affiliation{Instituto  de F\'{i}sica  Corpuscular, CSIC-UVEG,  Valencia 46071
  Spain} 
\author{Aldo M. Serenelli\thanks{E-mail: aldos@mpa-garching.mpg.de}}
\affiliation{Max-Planck-Institut f\"ur Astrophysik, Karl-Schwarzschild-Str. 1,
85748 Garching, Germany}

\begin{abstract}
Standard solar models (SSM) are facing nowadays a new puzzle: the solar 
composition problem.
New determinations of solar  metal abundances lead
SSM  calculations  to   conflict  with  helioseismological
measurements, showing  discrepancies that extend  from the convection  zone to
the solar core and can not be easily assigned to deficiencies in the modelling
of the solar convection zone. We present updated solar neutrino fluxes and
uncertainties  for two SSM  with high  (old) and  low (new)
solar  metallicity  determinations.  The  uncertainties  in  iron and  carbon
abundances  are the  largest contribution  to the  uncertainties of  the solar
neutrino fluxes.  The uncertainty  on the $^{14}$N(p,$\gamma$)$^{15}$O rate is
the  largest  of  the   non-composition  uncertainties  to  the  CNO  neutrino
fluxes. We  propose an independent  method to help  identify which SSM  is the
correct one.   Present neutrino  data can not  distinguish the  solar neutrino
predictions  of both models  but ongoing  measurements can  help to  solve the
puzzle.  
\end{abstract}

\pacs{Valid PACS appear here} \maketitle

\maketitle

This paper  is part of a  series led by John  Bahcall that spans  more than 40
years~\cite{series}.   The   goals  of  this  series  have   been  to  provide
increasingly  more  precise theoretical  calculations  of  the solar  neutrino
fluxes  and  detection  rates  and  to make  increasingly  more  comprehensive
evaluations of the uncertainties in the predictions.  

In  the  past,  standard  solar  models  (SSM) had  been  steadily  showing  a
discrepancy 
with   neutrino  measurements,   named  the   solar  neutrino   problem,  that
definitively led to the 
discovery of flavor change in the lepton sector. The significant difference of
the charged  current, electron scattering and neutral  current measurements of
the   epochal  Sudbury   Neutrino  Observatory   (SNO)   and  Super-Kamiokande
experiments  have demonstrated  the  new  phenomenon.  As  soon  as the  solar
neutrino puzzle was solved, a new and different puzzle has emerged.

Recent  refined  determinations  of  the  abundances  of  heavy
elements on the solar surface have 
led to lower heavy-element abundances 
~\cite{newcomp} than previously measured.  Solar models that incorporate these
lower abundances conflict 
dramatically  with helioseismological  measurements. Detailed  studies  of the
physical inputs have been done to resolve this controversy; 
with particular focus 
on  the region  where  more dramatically  low metallicity  solar
models 
fail to explain helioseismological measurements, 
right below  the solar convective envelope. But, recent
analysis of precise low-degree acoustic 
oscillations of the Sun  measured by the Birmingham Solar-Oscillations Network
(BiSON) indicate that the  discrepancies between solar models constructed with
low metallicity and helioseismic observations extend to the solar core and
thus  cannot  be attributed  to  deficiencies in  the  modelling  of the  solar
convection zone~\cite{chaplin}.  

A number of  studies have been done to refine  the estimated uncertainties and
search  for systematic  effects that  could  account for  the disagreement  of
helioseismological predictions of the low metallicity solar models.  Today, we
lack a solution to this puzzle and extra handles are needed to identify the
source  of the  discrepancy.   In this  letter,  we propose  a  new method  to
discriminate whether the low  heavy-element abundances should be accepted with
the other solar inputs in the SSM or not.  

Here we present  the most updated solar model calculations  with the two solar
abundance      determinations,      high      metallicity     labelled      as
BPS08(GS)~\cite{oldcomp}      and     low     metallicity      labelled     as
BPS08(AGS)~\cite{newcomp}.   There are  two steps  forward that  contribute to
major  improvements in  the estimated  precision  of the  neutrino fluxes:  i)
improved accuracy of the $^3$He-$^4$He cross section and ii) reduced systematic 
uncertainties in the determination of the surface composition of the Sun.

Previously, the uncertainty on the S-factor of the $^3$He($\alpha$,$\gamma$)$^7$Be
reaction  was  due  to an  average  discrepancy  in  results obtained  by  the
detection of the  delayed $\gamma$ rays from $^7$Be  decay and the measurement
of  prompt  $\gamma$ emission.  LUNA  collaboration  reported  on a  new  high
precision experiment using both techniques 
simultaneously at the
lowest interaction energy ever reached in the lab. The S-factors from the two
methods do not show any  discrepancy within the experimental errors, obtaining
S(0)=0.567$\pm$0.018$\pm$0.004  keV  barn~\cite{luna_S34}.   With  this  new
determination, the  uncertainty on the predicted $^8$B  ($^7$Be) neutrino flux
due  to  $S_{34}$  is reduced  from  7.5\%  to  2.7\%  (8.0\% to  2.8\%). 
Best-estimates   and   1-$\sigma$   uncertainties   of  the   other   important
astrophysical factors are summarized  in Table 1 of Ref.~\cite{10000ssm}, with
minor modifications~\cite{luna_S114}.  

Previously,   we  preferred   solar  model   calculations   with  conservative
uncertainties in  the element abundances, estimated by  the difference between
the high and low metal abundances, although it was recognized that estimated
uncertainties     based     on     each    analysis     independently     were
smaller~\cite{cons}.   New  detailed   studies   of  helioseismological   data
~\cite{chaplin} have shown that  low-degree helioseismology data clearly allow
to discriminate between predictions for the solar core structure from solar
models with high  and low metallicity. High
metallicity models agree with helioseismological data but not with
recent determinations  of solar abundances.  On the contrary,  low metallicity
models  are  inconsistent with  helioseismological  data  but  use the  recent
determinations of solar abundances. That is, high and low metallicity solar
models really are two different  groups of models, and conservative
uncertainties defined as described above are, in fact, too conservative. 
We find that uncertainties have  to be estimated for each class independently.
In what follows, we adopt for each solar abundances set~\cite{oldcomp,newcomp}
both the recommended values and uncertainties for each chemical element. 

\squeezetable
\begin{table}[!t]
\centering \caption[]{Predicted solar neutrino fluxes from solar
models. 
The table presents the predicted fluxes, in units of
$10^{10}(pp)$, $10^{9}({\rm \, ^7Be})$, $10^{8}(pep, {\rm ^{13}N,
^{15}O})$, $10^{6} ({\rm \, ^8B, ^{17}F})$, and $10^{3}(hep)$
${\rm  cm^{-2}s^{-1}}$.  Columns  2  and   3  show  BPS08  for  high  and  low
metalicities; and column 4 the flux differences between the models.  
\protect\label{tab:neutrinofluxes}}
\begin{tabular}{lccc}
\noalign{\bigskip} \hline\hline \noalign{\smallskip}
Source&\multicolumn{1}{c}{BPS08(GS)}&{BPS08(AGS)}& Difference\\
\noalign{\smallskip} \hline \noalign{\smallskip}
$pp$&5.97$(1 \pm 0.006)$&6.04$(1 \pm 0.005)$&1.2\%\\
$pep$&1.41$(1 \pm 0.011) $&1.45$(1 \pm 0.010) $&2.8\%\\
$hep$&$7.90 (1 \pm 0.15)$&$8.22 (1 \pm 0.15)$&4.1\%\\
${\rm ^7Be}$&$5.07 (1 \pm 0.06)$&$4.55 (1 \pm 0.06)$&10\%\\
${\rm ^8B}$&5.94$((1 \pm 0.11)$&4.72$(1 \pm 0.11)$&21\%\\
${\rm ^{13}N}$&$2.88 (1\pm 0.15) $&$1.89 (1~^{+0.14}_{-0.13}) $&34\%\\
${\rm ^{15}O}$&$2.15 (1~^{+0.17}_{-0.16}) $&$1.34 (1~^{+0.16}_{-0.15}) $&31\%\\
${\rm ^{17}F}$&$5.82 (1~^{+0.19}_{-0.17}) $&$3.25 (1~^{+0.16}_{-0.15}) $&44\%\\
\noalign{\smallskip} \hline \noalign{\smallskip}
Cl&$8.46 ^{+0.87}_{-0.88}$&$6.86 ^{+0.69}_{-0.70}$\\
Ga&$127.9 ^{+8.1}_{-8.2}$&$120.5^{+6.9}_{-7.1}$\\
\noalign{\smallskip} \hline\hline \noalign{\smallskip}
 \noalign{\smallskip}
\end{tabular}
\end{table}

Table~\ref{tab:neutrinofluxes} presents, in the second and third
columns, labelled BPS08(GS) and BPS08(AGS), our best solar model 
calculations for the neutrino fluxes. Both models were calculated with the new
input data for nuclear physics~\cite{luna_S34,luna_S114}. 
BPS08(GS)  was  computed with  the  high (old) metallicity   solar  abundances
determinations~\cite{oldcomp} and BPS08(AGS) with  the most recent analyses of
solar abundances~\cite{newcomp} that lead to conflicts with helioseismological
measurements.

Table~\ref{tab:neutrinofluxes}  also shows  the overall  uncertainties  in the
neutrino fluxes.   Uncertainties from  different sources have  been calculated
using     power-law     expansions      with     coefficients     given     in
Table~\ref{tab:powerlaw1}  and  added  quadratically.
Uncertainties due to 
errors  in  the solar  composition  (metals)  have  been determined  for  each
relevant metal separately~\cite{cons} and abundance uncertainties are
taken  from the  respective solar  composition  compilations~\cite{webpage}.  
This  is a  more consistent way for 
estimating fluxes  uncertainties and  leads to, incidentally,  reduced overall
uncertainties.  This  result is  particularly relevant in  the context  of CNO
fluxes  and  the solar  abundance  problem.  Solar  composition dominates  the
uncertainty in the  CNO fluxes (13\% for $^{13}$N, 12\%  for $^{15}$O and 17\%
for  $^{17}$F)  and  it is  not  dominant  for  all  other fluxes.  Among  the
non-composition sources, the most important contribution to the total error in
the CNO fluxes is due to the $S_{1,14}$ uncertainty.  
$S_{1,14}$ contributes
with 6\% (8\%),  while diffusion and opacity contribute with  4\% and 5\% (5\%
and 6\%) to the $^{13}$N ($^{15}$O) neutrino flux. For the $^7$Be flux, opacity
(3.2\%) contributes most to the total error, followed by
$S_{34}$ and  $S_{33}$ (2.8\%  and 2.5\%); while  for $^8$B,  opacity (6.8\%),
diffusion (4.2\%) and $S_{17}$ (3.8\%) dominate the uncertainties.

A  careful look to  the logarithmic  partial derivatives  with respect  to the
individual  solar  abundances  and  the uncertainties  of  the  individual
elements  shows  that the  major  contribution  to  the uncertainties  in  the
$^{13}$N and  $^{15}$O neutrino fluxes  largely comes from the  uncertainty in
the  carbon  abundance.   Therefore,  a  reduction  of  the  carbon  abundance
uncertainty  directly translates  into a  substantial improvement  of  the CNO
neutrino flux  uncertainties. On the other  hand, $^8$B and  $^7$Be fluxes are
much more sensitive to the iron abundance (due to its influence on the core
opacity) and therefore a reduction of  the iron
abundance uncertainty  directly translates  into a substantial  improvement of
the $^8$B and $^7$Be neutrino flux uncertainties.  

Finally, in
Table~\ref{tab:powerlaw1}  we  also included  power-law  coefficients for  the
dependence  on two  important  helioseismic quantities  on  solar model  input
parameters,  which   show  the  relevant  inputs  that   contribute  to  their
uncertainties.  SSM predictions for these quantities are $R_{CZ}=0.713 (0.728)
\pm 0.004$  and $Y_S=0.243 (0.229) \pm  0.035$ for the  high (low) metallicity
models;  while helioseismologically  determined values  are  $R_{CZ}=0.713 \pm
0.001$ and $Y_S=0.2485 \pm 0.0034$ \cite{helio}.

\squeezetable
\begin{table*}[htdp]
\centering   \caption[]{Logarithmic  partial   derivatives   $\alpha(i,j)$  of
  neutrino fluxes  and helioseismic quantities  with respect to  nuclear cross
  sections,  solar  luminosity  and  age, element  diffusion  rate,  radiative
  opacity and solar abundances.
\protect\label{tab:powerlaw1}}
\begin{tabular}{lcccccccccccccccccccc}
\noalign{\bigskip} \hline\hline \noalign{\smallskip}
Source & S$_{11}$ & S$_{33}$ & S$_{34}$ & S$_{17}$ & S$_{\rm hep}$ &
S$_{1,14}$ & S$_{^7Be,e}$ & L$_\odot$  & Age & Diff & Opac & C &  N & O & Ne &
Mg & Si & S & Ar & Fe \\
\noalign{\smallskip} \hline \noalign{\smallskip}
$pp$ &  0.090 & 0.029  & -0.059  & 0.000 &  0.000 & -0.004  & 0.000 &  0.808 &
-0.067 & -0.011 & -0.099 & -0.005 & -0.001 & -0.005 & -0.004 & -0.004 & -0.009
& -0.006 & -0.001 & -0.016 \\
$pep$ &  -0.236 & 0.043 & -0.086  & 0.000 & 0.000  & -0.007 & 0.000  & 1.041 &
0.017 & -0.016 & -0.300 & -0.009 &  -0.002 & -0.006 & -0.003 & -0.002 & -0.012
& -0.014 & -0.003 & -0.054 \\
$hep$ & -0.112  & -0.459 & -0.072 & 0.000  & 1.000 & -0.004 &  0.000 & 0.174 &
-0.118 & -0.037 & -0.398 & -0.007 & -0.002 & -0.020 & -0.014 & -0.017 & -0.036
& -0.028 & -0.005 & -0.064 \\
$^7Be$ &  -1.07 & -0.441 & 0.878  & 0.000 & 0.000  & -0.001 & 1.000  & 3.558 &
0.786 & 0.136 & 1.267 & 0.004 & 0.002  & 0.053 & 0.044 & 0.057 & 0.116 & 0.083
& 0.014 & 0.217 \\ 
$^8$B & -2.73 & -0.427 & 0.846 & 1.000 & 0.000 & 0.005 & 0.000 & 7.130 & 1.380
& 0.280  & 2.702 & 0.025  & 0.007 &  0.111 & 0.083 &  0.106 & 0.211 &  0.151 &
0.027 & 0.510 \\
$^{13}N$ &-2.09  & 0.025 & -0.053  & 0.000 & 0.000  & 0.711 & 0.000  & 4.400 &
0.855 & 0.340 & 1.433 & 0.861 & 0.148  & 0.047 & 0.035 & 0.051 & 0.109 & 0.083
& 0.015 & 0.262 \\
$^{15}O$ & -2.95  & 0.018 & -0.041 & 0.000  & 0.000 & 1.000 &  0.000 & 6.005 &
1.338 & 0.394 & 2.060 & 0.810 & 0.207  & 0.075 & 0.055 & 0.076 & 0.158 & 0.117
& 0.021 & 0.386 \\
$^{17}F$ & -3.14  & 0.015 & -0.037 & 0.000  & 0.000 & 0.005 &  0.000 & 6.510 &
1.451 & 0.417 & 2.270 & 0.024 & 0.005  & 1.083 & 0.061 & 0.084 & 0.174 & 0.128
& 0.023 & 0.428 \\
\noalign{\smallskip} \hline
$R_{CZ}$ & -0.061 & 0.002 & -0.003 & 0.000 & 0.000 & 0.000 & 0.000 & -0.016 & 
-0.081 & -0.018 & -0.012 & -0.006 &  -0.005 & -0.028 & -0.012 & -0.005 & 0.002
& 0.004 & 0.001 & -0.009 \\ 
$Y_S$ & 0.134 & -0.005 & 0.009 & 0.000 & 0.000 & 0.001 & 0.000 & 0.373 & 
-0.110 & -0.073  & 0.646 & -0.009 & -0.001  & 0.023 & 0.033 &  0.037 & 0.070 &
0.048 & 0.009 & 0.089 \\
\noalign{\smallskip} \hline \hline \noalign{\smallskip}
\noalign{\smallskip}
\end{tabular}
\end{table*}

Opacity  uncertainties  have  been  conservatively computed  by  defining  the
1-$\sigma$ uncertainty as given by  the flux differences between two SSMs that
have been computed one with standard opacities and the other with opacities in
the solar interior increased by 2.5\%.  Such a difference in the solar core
opacities is  representative of  the maximum difference  between the  two most
up-to-date radiative opacitity calculations, OPAL and OP~\cite{op-opal}. 

Based on the  results presented above, we make  two recommendations for future
work  to  improve the  uncertainties  in  the  SSM  neutrino
fluxes. First, the uncertainty in the  low energy extrapolation of the rate of
the  $^{14}$N(p,$\gamma$)$^{15}$O  reaction should  be  reduced  to below  5\%
(almost a 
factor 2 improvement) in order that the uncertainty due to astrophysical cross
sections does  not dominate  the non composition  uncertainties in any  of the
neutrino  fluxes.   Second, the  uncertainty  in  the surface  element
abundances iron and carbon should be reduced to $\pm$ 0.02  dex 
(factor 2 to 3 improvement) to significantly improve the radiative opacity and 
the CNO  composition uncertainties. The requirements are tough, but the outcome 
is rewarding. 
 
Table~\ref{tab:neutrinofluxes} shows  that the central values  of the neutrino
fluxes in  the two  models BPS08(GS)  and BPS08(AGS) differ  by 2-3  times the
theoretical error expected within one  model. This feature must be explored to
estimate by how much solar neutrino experiments can contribute to solve 
the solar  composition puzzle.  Two of the  neutrino fluxes ($^8$B  and $^7$Be
neutrinos) 
have been measured with good precision, and this can lead to preliminary tests
of the solar composition.  
SNO collaboration \cite{sno1,sno2,sno3} has determined  the $^8$B neutrino flux with good
precision by the measurement of the charged current and the neutral current 
detection of solar neutrinos in three phases of the experiment: I) heavy water
target, II) heavy  water + salt target and III) heavy  water + $^3$He targets.
All three 
phases lead to a very precise measurement of the $^8$B flux ($\sim$ 6\%). 
Moreover, Super-Kamiokande electron scattering data, when combined with
the other  solar neutrino  data and the  assumption of  neutrino oscillations,
contributes to 
further reduce the uncertainty in the $^8$B flux. 
Since May 2007, the Borexino detector is collecting data on the electron
scattering of $^7$Be neutrinos.  The collaboration presented the latest 
results on  $^7$Be solar  neutrinos ($\sim$10\% precision)  based on  192 live
days  of  data  taking  \cite{borexino}.   With  more  statistics  and  better
calibrations, the collaboration anticipates a final precision at the 5\% level. 

In order to quantitatively determine how well the two solar models fit 
the neutrino data we proceed in two steps: a) use all of the available solar 
and reactor neutrino data to determine the current constraints on neutrino 
oscillation parameters and solar neutrino fluxes with the method described in 
Ref.~\cite{roadmap}, b) use the solar neutrino fluxes inferred from data to 
test SSM fluxes. We discuss now which data have been included 
in the analysis and refer to Ref.~\cite{roadmap} for further details. Solar 
neutrino data includes the average rate measured by the radiochemical 
experiments, chlorine~\cite{chlorine} and gallium~\cite{gallium}, the 
SuperKamiokande I zenith spectrum~\cite{sk1}, the SuperKamiokande II day-night
spectrum~\cite{sk2}, the SNO pure D$_2$O phase day-night spectrum~\cite{sno1},
the SNO  salt phase  day-night spectrum~\cite{sno2}, the  electron scattering,
charged current and  neutral current rates measured in  phase 3~\cite{sno3} and
the Borexino  $^7$Be and $^8$B rates~\cite{borexino}. Non  solar neutrino data
include the antineutrino data from KamLAND~\cite{kamland} and the marginalized
$\chi^2(\theta_{13})$ function derived from the analysis of 
atmospheric+K2K+MINOS+CHOOZ data ~\cite{theta13}.  The free parameters in the 
analysis are the neutrino parameters $\Delta m^2_{21}$, $\theta_{12}$ and 
$\theta_{13}$  and the  reduced neutrino  fluxes $f_{\mathrm  B}$, $f_{\mathrm
  Be}$, $f_{\mathrm pp}$, 
$f_{\mathrm  N}$  and $f_{\mathrm  O}$  ($f_i$  is  the $i$-th  neutrino  flux
normalized to the 
BPS08(GS) prediction) subject to the luminosity constraint, i.e., the total 
luminosity produced by nuclear reactions (the p-p chains and the CNO cycle) 
equals the observed solar luminosity.  We find that: 
\begin{eqnarray*}
\begin{array}{rclcrcl}
\Delta m^2_{21}&=&\multicolumn{4}{l}{(7.7 \pm 0.2)\times 10^{-5} {\mathrm eV}^2}\\
\tan^2       \theta_{12}&=&0.46^{+0.04}_{-0.05}        &       ;       &\sin^2
\theta_{13}&=&0.014^{+0.011}_{-0.009} \\ 
f_{\mathrm B}&=&0.91 \pm 0.03 &;& f_{\mathrm Be}&=&1.02 \pm 0.10 \\
f_{\mathrm pp}&=&1.00^{+0.01}_{-0.02} &;& L_{\mathrm CNO}&=&0.0^{+2.9}_{-0.0} \%
\end{array}
\end{eqnarray*}
where $L_{\mathrm  CNO}$ is the  fraction of solar  luminosity due to  the CNO
nuclear 
fusion reactions.  The oscillation parameters, central values  and errors, are
in excellent 
agreement with neutrino analysis where the solar neutrino fluxes are not
treated as free variables \cite{theta13,lisi}.
Compared to previous 
analysis \cite{roadmap}, the major improvement  is in the determination of the
$^7$Be flux. The Borexino measurements 
lead to a  factor five improvement of the uncertainty in  this flux. The upper
bound on the CNO luminosity is not significantly different \cite{cno} 
because the  improvement due to the  $^7$Be determination leading  to the more
constrained  $^{13}$N and $^{15}$O  fluxes, compensates  the lower  CNO fluxes
predicted  by the  solar  model  calculations using  the  most accurate  cross
section of the  $^{14}$N(p,$\gamma$)$^{15}$O reaction \cite{luna_S114} used in
BPS08.  The results of this  analysis allow us to experimentally determine the
ratio    of     the    termination     chains    in    the     p-p    burning:
R=$\langle^3$He+$^4$He$\rangle$/$\langle^3$He+$^3$He$\rangle$=0.19$\pm$0.02.  

The  global   analysis  of   neutrino  data  used   above  was   discussed  in
Refs.~\cite{roadmap,cno} with 
the goal to derive neutrino oscillation parameters and solar neutrinos fluxes 
from neutrino data independently of the solar model. 
We can now move to the second step in the analysis and 
test SSM predictions without  dealing with solar neutrino data, which is  
encoded in  the results presented above.   In order to  statistically test the
hypothesis that  there is no difference between a given  solar model and the
data, we define the $\chi^2$ function 
\begin{eqnarray*}
\chi^2 &=& \sum_{ij} (f^{th}_i-f_i) \sigma^{-2}_{ij}(f^{th}_j-f_j) \\
\sigma^2_{ij}                                                               &=&
\sigma_{{\mathrm         exp},i}\sigma_{{\mathrm         exp},j}\rho^{{\mathrm
    exp}}_{ij}+\sigma_{{\mathrm   th},i}\sigma_{{\mathrm  th},j}\rho^{{\mathrm
    th}}_{ij} 
\end{eqnarray*}
where   $\sigma_{exp,i}$  and   $\sigma_{th,i}$  are   the   experimental  and
theoretical errors of the  neutrino flux $f_i$ and $\rho^{{\mathrm exp}}_{ij}$
and 
$\rho^{{\mathrm th}}_{ij}$  are the experimental  and theoretical correlations
of the 
fluxes i  and j. 
Given the  large uncertainties in some  of the experimentally
determined fluxes, we only include in the analysis the most precise ones
($^8$B    and   $^7$Be   fluxes),    which   have    negligible   experimental
correlations. Theoretical correlation coefficients  for neutrino fluxes can be
found in Ref.~\cite{webpage}. 
The analysis leads to:
\begin{eqnarray*}
\begin{array}{lcl}
\chi^2_{\mathrm BPS08(GS)} = 0.9~(63\%) & {\rm and}&
\chi^2_{\mathrm BPS08(AGS)} = 1.5(47\%),
\end{array}
\end{eqnarray*}
where the goodness of fit is shown  in parenthesis and is used as a diagnostic
for the fit of the model 
to the data. Neutrino  data  are  not precise  enough  
to reject the null hypothesis for any of the models, BPS08(GS) and BPS08(AGS).
Adding the pp flux in the analysis does not improve the
statistical  power, $\chi^2$ changes  to 0.95  and 1.6  respectively. These
results show that  the theoretical and statistical errors  are still too large
to 
reject any of the  two models. The analysis  proposed here will guide us on how
improvements can contribute to solve the composition problem. Eventually, the 
rejection of  the null hypothesis  would prompt us  to use a  likelihood ratio
testing for a 
given parameter(s)  connecting the models. More Super-Kamiokande  data and the
full analysis 
of the SNO data, with lower energy threshold, will sharpen the accuracy of the 
$^8$B  neutrino   flux.  Borexino  ongoing   measurements  anticipate  further
improvements of the 
statistical and  systematic errors which  may help to discriminate  models. To
illustrate what 
might  be achieved,  if  we assume  that  future neutrino  data  will lead  to
$f_{\mathrm B}=1.0 \pm 0.02$ 
and $f_{\mathrm  Be}=1.00 \pm 0.05$, and  theoretical errors on  the $^8$B and
$^7$Be fluxes are 
reduced  from  $\pm  0.11$  and  $\pm  0.06$ to  $\pm  0.08$  and  $\pm  0.04$
respectively, we obtain 
$\chi^2_{\mathrm  BPS08(GS)}  =  0$  (by  construction)  and  $\chi^2_{\mathrm
  BPS08(AGS)} = 6.2~(4.5\%)$, 
what will hint towards the relevance of doing an analysis of solar models as a
function of the iron 
composition to  all neutrino  data. Moreover, running  and planned  low energy
experiments have a realistic 
chance  to   measure  CNO  neutrinos  and  further   distinguish  the  generic
predictions of both models, 
given that predictions of BPS08(GS) and BPS08(AGS) for these fluxes differ about
30\%. This kind of study, however, is beyond the  scope of this
paper. The hypothesis testing 
analysis  with  future data  will  show whether  this  effort  is valuable  to
discriminate between the high and low composition SSM.

In summary, standard solar  models are facing the solar composition problem,
i.e., new composition determinations,  when included in solar models, conflict
with  helioseismological  measurements.   We  are  agnostic  whether  the  new
composition,   the  solar  modelling,  other   inputs  or  a
combination  of them  is the  source  of the  conflict.  We  have proposed  an
independent method to  identify which solar model is the  correct one by using
 solar  neutrino data.  A method  of analysis was  proposed and preliminary
results have  been derived. Ongoing  solar neutrino measurements  and forecast
neutrino flux  measurements will help  in the search  of the solution  to this
puzzle.

\acknowledgments We are deeply indebted to the late John Bahcall for his 
forty years of continuous work on refining the solar model calculations and 
testing predictions  with all available neutrino  data.  We thank  E. Lisi for
useful suggestions and for careful reading of the manuscript. We thank PHYSUN 
organizers and participants  for the encouragement to complete  this work. CPG
acknowledges MICINN grant No. FPA2007-60323 and thanks hospitality at TMU 
where this manuscript was completed.

\end{document}